\documentclass[twocolumn,prl,aps,epsf,floats]{revtex4}

	\newif\ifpdf
	\ifx\pdfoutput\undefined
	\pdffalse 
	\else
	\pdfoutput=1 
	\pdftrue
	\fi

	\ifpdf
	\usepackage[pdftex]{graphicx}
	\else
	\usepackage{graphicx}
	\fi

\begin{document}

 \ifpdf
	\DeclareGraphicsExtensions{.pdf, .jpg, .tif}
	\else
	\DeclareGraphicsExtensions{.eps, .jpg}
	\fi

\title{Wigner crystallization about $\nu=3$}

\author{ R.\ M.\ Lewis$^{1,2,}$\footnote{ corresponding author, email: rlewis@magnet.fsu.edu, fax: 850 644-5038}, Yong Chen$^{1,2}$, L.\ W.\ Engel$^{1}$, D.\ C.\ Tsui$^{2}$, \\
P.\ D.\ Ye$^{1,2}$, L.\ N.\ Pfeiffer$^{3}$, and K.\ W.\ West$^{3}$
}

\address{$^{1}$NHMFL,1800 E. Paul Dirac Dr., Florida State University, Tallahassee, FL 32310, USA\\
$^{2}$Dept.\ of Electrical Engineering, Princeton University, Princeton, NJ 08544\\
$^{3}$Bell Laboratories, Lucent Technology, Murray Hill, NJ 07974}

\begin{abstract}
We measure a resonance in the frequency dependence of the real diagonal conductivity, Re[$\sigma_{xx}$], near integer filling factor, $\nu=3$.  This resonance  depends strongly on $\nu$, with peak frequency $f_{pk} \approx 1.7$ GHz  at $\nu=3.04$ or 2.92 close to integer $\nu$, but $f_{pk} \approx$ 600 MHz at $\nu=3.19$ or 2.82, the extremes of where the resonance is visible. 
 The dependence of $f_{pk}$ upon $n^*$, the density of electrons in the partially filled level, is discussed and compared with similar measurments by Chen {\it et al.}\cite{yong} about $\nu=1$ and 2.  We interpret the resonance as due to a pinned Wigner crystal phase with density $n^*$ about the $\nu=3$ state. 
\end{abstract}

\maketitle    


The integer quantum Hall effect (IQHE) \cite{vklitzing,reviews} is understood to arise in two--dimensional electron systems (2DES) from gaps in the single particle density of states.  At low temperatures, electron localization when the Fermi level resides in these gaps, gives rise to wide, deep minima in the diagonal conductivity, $\sigma_{xx}$, and quantized values of the Hall conductance, $\sigma_{xy}$.    In very clean systems, there exists the possibility that electrons may become localized in a collective manner forming crystalline domains\cite{MacDgirvin,fogler,yoshioka,kun} pinned by disorder.  Collective localization of electrons is known to occur deep in the lowest Landau level at filling factors $\nu < 1/5$ where the Wigner crystal (WC)\cite{WCreviews} becomes energetically favorable \cite{lamgirvin,levesque}.  Key observations of the WC have been provided by measurements of the real diagonal microwave conductivity, Re[$\sigma_{xx}$], which shows a resonance at frequency, $f \sim 1$ GHz \cite{lloydssc,peide}.  The resonance is caused by the pinned WC domains oscillating in the disorder potential, i.e. a pinning mode\cite{flprb,normand}.  Recently, Chen {\it et al.} \cite{yong} have shown that a similar resonance exists at $\nu$ close to integers 1 and 2 and attributed it to a crystal formed from the electrons in the uppermost occupied spin split Landau level.


In this paper, we detail measurements of a resonance in Re[$\sigma_{xx}]$ about $\nu=3$.  We interpret this resonance as due to a pinned crystal phase about $\nu=3$.  This resonance is sharp and has strong $B$ dependence in agreement with Ref. \cite{yong}.  The dependence of $f_{pk}$ upon the partial density in the uppermost occupied spin split Landau level is shown to agree well with measurements of $f_{pk}$ about $\nu=1$ and 2.  It also shows that Landau level index is important in determining the properties of  integer quantum Hall Wigner crystals  (IQHWC).  

Our measurements were performed on an MBE grown GaAs/AlGaAs 300{\AA} quantum well of density $n=3.03 \times 10^{11} \ {\rm cm}^{-2}$ and mobility $\mu =2.4 \times 10^7 \  {\rm cm^{2} \  V^{-1}\  s^{-1}} $.  The sample was cleaved to approximately $ 4 \times 6$ mm. A coplanar waveguide (CPW) was patterned onto the sample surface such that microwave signals propagate across the sample with the microwave electric field in the plane of the 2DES.  In the high $B$ limit, the absorbed power gives the Re[$\sigma_{xx}$] = $ \frac{-w}{2lZ_0} \ln(P/P_0)$ where $P$ is the transmitted power, $P_0$ is the power transmitted in the absence of a 2DES, $Z_0= 50 \Omega$, and $\frac {w}{2l}$ is a geometric factor equal to the inverse number of squares of 2DES in the CPW\cite{lloydprl93}.  The sample is mounted on a metal block which itself is well heat sunk to the mixing chamber and is positioned in the superconducting magnet.  The temperature, $T$, of the block is measured by a resistor.

\begin{figure}[tb]
\begin{center}
\includegraphics[width=8cm]{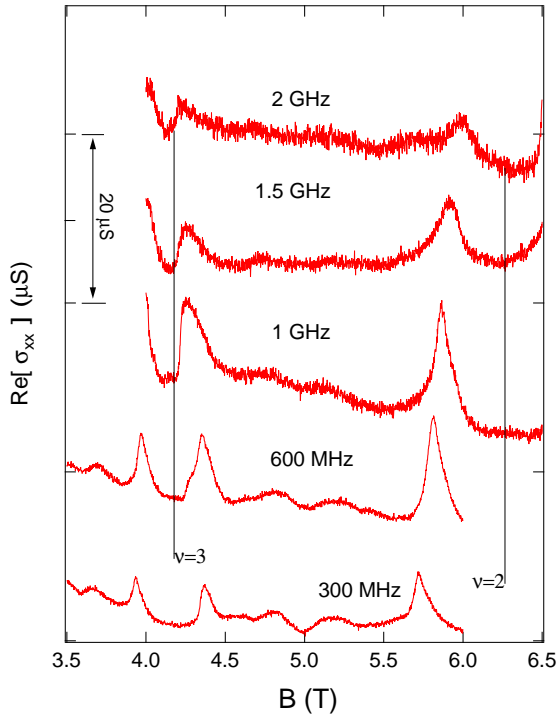}
\caption{ The real part of the conductivity, Re[$\sigma_{xx}$] versus $B$ between 3.5 and 6.5 T at $T \approx$ 50 mK. Five frequencies, .3, .6, 1.0, 1.5, and 2 GHz are shown.  The data are offset for clarity.}
\label{fig.1}
\end{center}
\end{figure}

In figure 1, we plot Re[$\sigma_{xx}$] between 3.5 and 6.5 T at 5 frequencies.  These data were acquired at $T \approx 50$ mK and are vertically offset for clarity.  The 300 MHz data show a wide, broad minimum centered about $B=4.18$ T, which is $\nu=3$.  A second minimum begins to develop as $B$ approaches 6.26 T, which is $\nu=2$.  Sharp frequency dependent minima are also present at 5.01 T and 3.58 T, which are $\nu=5/2 \ {\rm and } \ 7/2 $ respectively.  Focusing on the region around $\nu=3$, the peaks on each side of the minimum at $B=4.37 $ and 3.93 T are about 5 $\mu$S in size.  In the 600 MHz data, the peaks have grown to about 8 $\mu$S and have also moved inwards slightly to $B=3.97$ and 4.35 T.  Data at 1 GHz show the high $ B$ side of $\nu=3$ peak even larger at 9 $\mu$S and  $B=4.26$ T.  By 1.5 GHz, this peak has shrunk to 5 $\mu$S, and by 2 GHz, the peak is just 4 $\mu$S.  However, as $f$ increases, it continues to shift inward towards $\nu=3$ peaking at $B=4.25$ and 4.21 T respectively.  Similar behavior can be seen on the low B side of $\nu=2$ where the peak reaches its maximum conductivity at about 1 GHz and shrinks thereafter with increasing $f$, while shifting continuously towards $\nu=2$.  

\begin{figure}[tb]
\begin{center}
\includegraphics[width=8cm]{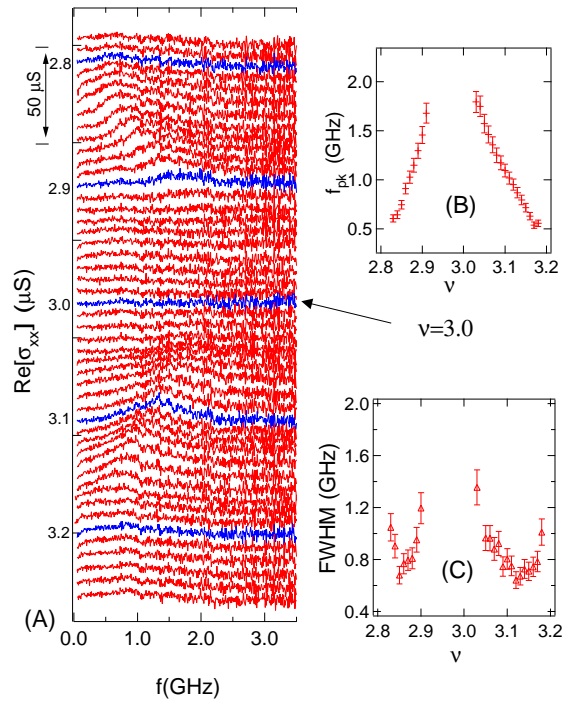}
\caption{{\bf (A)} The real part of the conductivity, Re[$\sigma_{xx}$] versus $f$ at filling factors, $\nu$, between 3.78 and 4.24.{\bf (B)} The peak frequency,  $f_{pk}$ versus $\nu$. {\bf (C)}  Full width at half maximum, FWHM, versus $\nu$.}
\label{fig.2}
\end{center}
\end{figure}
Figure 2 shows Re[$\sigma_{xx}$] versus $f$ at fixed $\nu$ about $\nu=3$ from 2.78 (top trace) to 3.24 (bottom trace).  The data with darker lines fall at $\nu=2.8, 2.9, 3.0, 3.1$, and 3.2 which are marked on the left and side.  All data were acquired at $ T  \sim 40$ mK and successive traces are offset in 6 $\mu$S increments.  A resonance is seen for $\nu$ between 2.80 and 2.92 on the low side of $\nu=3$ and again between $\nu=3.04$ and 3.19 on the high side.   In Fig. 2B, the peak frequency $f_{pk}$ of the resonance is plotted for those traces to which a lorentzian curve can reasonably be fitted.  The $f_{pk}$ runs from 600 MHz at $\nu=2.83$ to 1.7 GHz at $\nu=2.91$.  Above $\nu=3$, $f_{pk}$ fall between 1.8 GHz at $\nu=3.04$ and about 560 MHz at $\nu=3.18$.  Fig. 2C shows the full width at half maximum, FWHM, taken from the same lorentzian fit.  The resonance reaches its strongest (narrowest FWHM) at $\nu=2.86$ and 3.12.  In both cases, FWHM$ \approx 650$ MHz.  The errors on the FWHM are about 10 $\%$.  Those on $f_{pk}$ are about 6 $\%$ and the error in $\nu$ is less than $\pm 0.02$.

A resonance of this kind in Re[$\sigma_{xx}$], as has been discussed elsewhere\cite{rlewis2,yong}, is naturally interpreted as a pinning mode\cite{flprb} of an electronic crystal.  The resonance observed here varies from well above the temperature (40 mK) at 1.7 GHz to well below it at 650 MHz (1 GHz $\times {\rm h}/k_B \ \approx 50$ mK) and persists to temperature well in excess of 100 mK.  Hence, it is unlikely that individually trapped electrons are its cause.  The resonance is sufficiently sharp with $Q=f_{pk}/{\rm FWHM} >1$ that some averaging of pinning potentials is occurring, indicating a large domain size.  Finally, the data can be compared to the resonance seen in the Wigner Crystal regime, $\nu < 1/5$ \cite{peide}, to the similar resonance seen about $\nu=1$ and 2 \cite{yong}, and to the resonance\cite{rlewis2} of the bubble phase.

	The resonance behavior seen here and in ref. \cite{yong} is very different from what has been seen in the IQHE minima before.  In moderately clean\cite{note2} 2DES over a similar range of frequency, $1 \le f \le 10$ GHz, microwave conductivity experiments\cite{rlewis1,hohls} found Re$[\sigma_{xx}] \propto f$, within the QHE minima.  Those data were consistent with single particle hopping theory\cite{polyakovshklovkii}.  In contrast, the pinning resonance seen here is due to the collective motion of pinned crystalline domains.

 The dependence of the resonance on $\nu$ is mainly due to a change in the density of electrons/holes in the crystal.  Assuming that filled levels are inert, the remaining electrons/holes in the partially filled/empty level above/below $\nu=3$ have density, $n^* = n \times \nu^* /\nu$ where $\nu^*=\nu-$[$\nu$] is the partial filling of the level and in this case, [$\nu$]=3.

\begin{figure}[tb]
\begin{center}
\includegraphics[width=8cm]{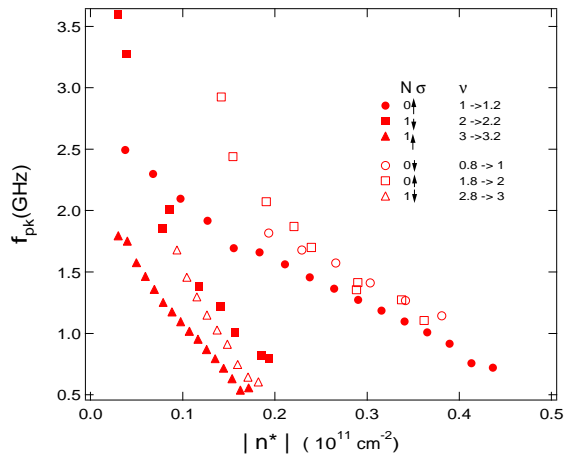}
\caption{ Peak frequency, $f_{pk}$ versus partial density, $|n^*|$, about integer filling, $n^*= n \times \nu^*/\nu$ . Filled symbols represent electrons and open symbols holes.  Data from the resonance about $\nu=1$ and 2 are included.  The quantum numbers  N and $\sigma$ are also given. }
\label{fig.3}
\end{center}
\end{figure}

In figure 3 we plot $f_{pk}$ versus $|n^*|$ for the resonance about $\nu=3$, 2, and 1\cite{comment1}.   Data for the resonance above integer $\nu$ (electrons) are shown as filled symbols and those from holes (just below integer $\nu$) are open symbols.  The quantum numbers for Landau level index, N, and spin, $\sigma$, are also given.  While $f_{pk}$ monotonically decreases as $n^*$ increases in all cases, the data naturally bunch into 2 groups.  Around $\nu=3$ (triangles),  the $f_{pk}$ for  electrons (1,$\uparrow$) and holes (1,$\downarrow$) fall close to each other.  Both data sets agree well with data from the electrons near $\nu = 2$ (filled squares) (1, $\downarrow$).  All three occur in the N=1 Landau level.   However, the other three data sets were taken at ranges of $\nu$ which fall in the lowest Landau level.  These are, near $\nu=1$,  electrons (0,$\uparrow$) and holes (0,$\downarrow$)  and holes below $\nu=2$ (0,$\downarrow$) the data for all of which follow a more gradual slope.  These two distinct bunchings are due to differences between in the single particle wave function in the N=0 and N=1 Landau levels.

In conclusion, we have presented data for an integer Wigner crystal phase of electrons whose density comes from the partial filling of the N=1 Landau level about $\nu=3$.  Our evidence is a strong resonance in Re$[\sigma_{xx}]$ at $\nu$ close to 3, with a range from 2.83 to 2.91 and again from 3.04 to 3.19.  The best $Q$ are measured at $\nu=2.86$ and 3.12.  The dependence of $f_{pk}$ upon $\nu^*$ closely resembles that of the previously measured resonance near $\nu=1$ and 2 and suggests that Landau Level index is of principal importance to the properties of the resonance.

We thank Kun Yang, Nick Bonsteel, and Ravin Bhatt for stimulating discussions.  This work is supported by a grants from the AFOSR and the NHMFL in house research program.


\begin{thebibliography}{9}
\bibitem{comment1}The $f_{pk}$ data for resonances around $\nu=1$ and 2 were measured in a separate cooldown on a different sample cut from the same wafer.
\end{thebibliography}
\end{document}